\UseRawInputEncoding
\documentclass[sn-mathphys,Numbered]{sn-jnl}

\usepackage{graphicx}%
\usepackage{multirow}%
\usepackage{amsmath,amssymb,amsfonts}%
\usepackage{amsthm}%
\usepackage{mathrsfs}%
\usepackage[title]{appendix}%
\usepackage{xcolor}%
\usepackage{textcomp}%
\usepackage{manyfoot}%
\usepackage{booktabs}%
\usepackage{algorithm}%
\usepackage{algorithmicx}%
\usepackage{algpseudocode}%
\usepackage{listings}%
\usepackage{graphicx}%
\usepackage{subcaption}%
\usepackage{amsthm}%
\usepackage{siunitx}

\theoremstyle{thmstyleone}%
\theoremstyle{thmstyletwo}%

\theoremstyle{thmstylethree}%

\begin{document}

\title[Article Title]{
Measuring the Kapitza Resistance between a Passivated Semiconductor and Liquid Helium}

\author*{\fnm{Babak} \sur{Mohammadian}*}\email{Babak.mohammadian@postgrad.manchester.ac.uk}
\author{\fnm{Mark A.} \sur{McCulloch}}

\author{\fnm{Valerio} \sur{Gilles}}
\author{\fnm{Thomas} \sur{Sweetnam}}
\author{\fnm{Lucio} \sur{Piccirillo}}

\affil{\orgdiv{Jodrell Bank Centre for Astrophysics}, \orgname{University of Manchester}, \orgaddress{\city{Manchester}, \country{UK}}}


\abstract{
In this paper, we describe an experimental investigation into the effect of passivation layer thickness on heat dissipation between a quartz substrate and liquid helium. We have observed that by depositing SiN from 0 to 240 nm, the Kapitza resistance increases by $\sim$ 0.0365 m$^2$.K/W  per nanometer more than for an unpassivated semiconductor. We hypothesize that this increase in Kapitza resistance represents an additional barrier to the cooling of semiconductor devices in liquid helium. }



\maketitle
 \vskip-2.5em
\section{Introduction}\label{sec1}
Several areas of fundamental and applied physics would benefit from amplifying weak microwave signals at or close to the quantum noise level. These include QUBITS (or quantum bit) in quantum computing systems, astronomical observations, and dark matter experiments \cite{1,2,LNAobserv,dark}. Most of these experiments currently use High Electron Mobility Transistors (HEMTs) based amplifiers. However, the typical noise temperature of the best amplifiers is of the order of 6-10 times the quantum noise limit (50 mK/GHz) \cite{bardin2017cryogenic}. This is attributed to the fact that once they are cooled below 20 K, the HEMT’s conduction channel remains hotter than its surroundings. This is believed to be because phonon radiation can no longer remove the heat generated by resistance in the conduction channel, resulting in self-heating that prevents the area around the conduction channel from cooling below 18 K \cite{8,9,10,11,12}. Self-heating has also been observed at room temperature in HEMTs used for power amplifiers \cite{14}. It has been reported that the thickness and composition of the passivation layer, a 200-700 nm coating of SiN that is applied to HEMTs to prevent degradation of the electrical performance, has an impact on the self-heating.\\
To try to overcome self-heating at cryogenic temperature, a recent experiment \cite{13} immersed a HEMT in normal (He I) and superfluid (He II) \textsuperscript{4}He baths. However, no further reductions in noise temperature were observed. We hypothesize that the SiN passivation layer that is typically applied to electronic devices, may be applying an additional barrier to the removal of heat.\\
At cryogenic temperature, the effect of surface passivation on heat dissipation by phonons has been evaluated in liquid nitrogen (77 K) by the authors \cite{babak}. In this earlier paper, we showed that the increase in the thickness of the passivation layer resulted in an increase in the thermal boundary (Kapitza) resistance at the interfacial area.\\  In this paper, we extend our investigation on surface passivation and Kapitza resistance to measurements in liquid helium (4.2 K). In section \ref{sec:II}, we discuss the theory of Kapitza resistance and how it is related to acoustic impedance mismatch. Then, sections \ref{sec:III} and \ref{sec:IV} present the experiment, including how we calibrate the capacitance thermometer and our Kapitza resistance measurements in the liquid helium. Finally, in section \ref{sec:V}, we discuss our results and how they could be applied to reducing the self-heating in transistors.
\section{Thermal Dissipation}\label{sec:II}
 The heat transfer across the interface between a solid material and liquid helium 
 results in a temperature difference between each side of the interface. Since the boundary between a liquid and a solid is atomically flat, a thermal boundary resistance, also known as Kapitza resistance, $R_K$, is given by the ratio of the temperature ($\Delta{T}$) across the interface and the heat flow ($\dot{Q}$) through the interface which defines a thermal resistance for this boundary \cite {b16}:
\begin{equation}
R_{K} = A \cdot \frac{{\Delta{T}}}{{\dot{Q}}}  
 (\mathrm{m}^2\mathrm{K/W})
\label{eq:1}
\end{equation}
where A is the area of the interface.\\
\noindent Kapitza resistance arises due to the mismatch in the acoustic impedance (the product of density and sound velocity) between a solid material and liquid helium \cite{khalta}, and can result in a considerable impedance to the transmission of phonons across the interface.\\ 
Table \ref{table:densityvelocityimpedance} provides data on the sound velocity and density of materials of interest where the acoustic impedance of different mediums can be compared. It can be seen that the acoustic impedance for SiN is significantly larger than for liquid helium. As a result of this mismatch, a proportion of phonons, incident on both sides of the interface, are reflected back. The values in Table \ref{table:densityvelocityimpedance} are based on room temperature due to available references. It is important to note that the sound velocity at 4 K would likely be slightly lower compared to its value at room temperature.\\  
\newpage 

\begin{table}[h!]
\centering
\caption{Density, velocity, and acoustic impedance values for different materials}
\begin{tabular}{c c c c} 
 \hline
Material & Density (kg/$m^3$) & Sound Velocity ($m/s$) & Acoustic Impedance (Pa$\cdot$ s/$m^3$) \\ [1ex]

 \hline\hline

  Au \cite{gold} & $19,320$ & $3.24 \times 10^3$ & $6.26 \times 10^7$ \\
SiN\cite{SiN} & $3,170$ & $1.12 \times 10^4$ & $3.56 \times 10^7$ \\
Silicon \cite{silicon}& 2,340 &$5.94 \times 10^3$&$1.39 \times 10^7$\\
 Fused Quartz\cite{Fusedqua} & $2,210$ & $5.88 \times 10^3$ & $1.30 \times 10^7$ \\
Liquid Helium\cite{liquidhe11} & $125$ & $189.2$ & $2.37 \times 10^4$ \\

 \hline
\end{tabular}
\label{table:densityvelocityimpedance}
\end{table}
\vspace{-0.3cm}
\section{The Experiment}\label{sec:III}
\begin{figure}[ht]
    \centering
           \includegraphics[width=0.68\linewidth]{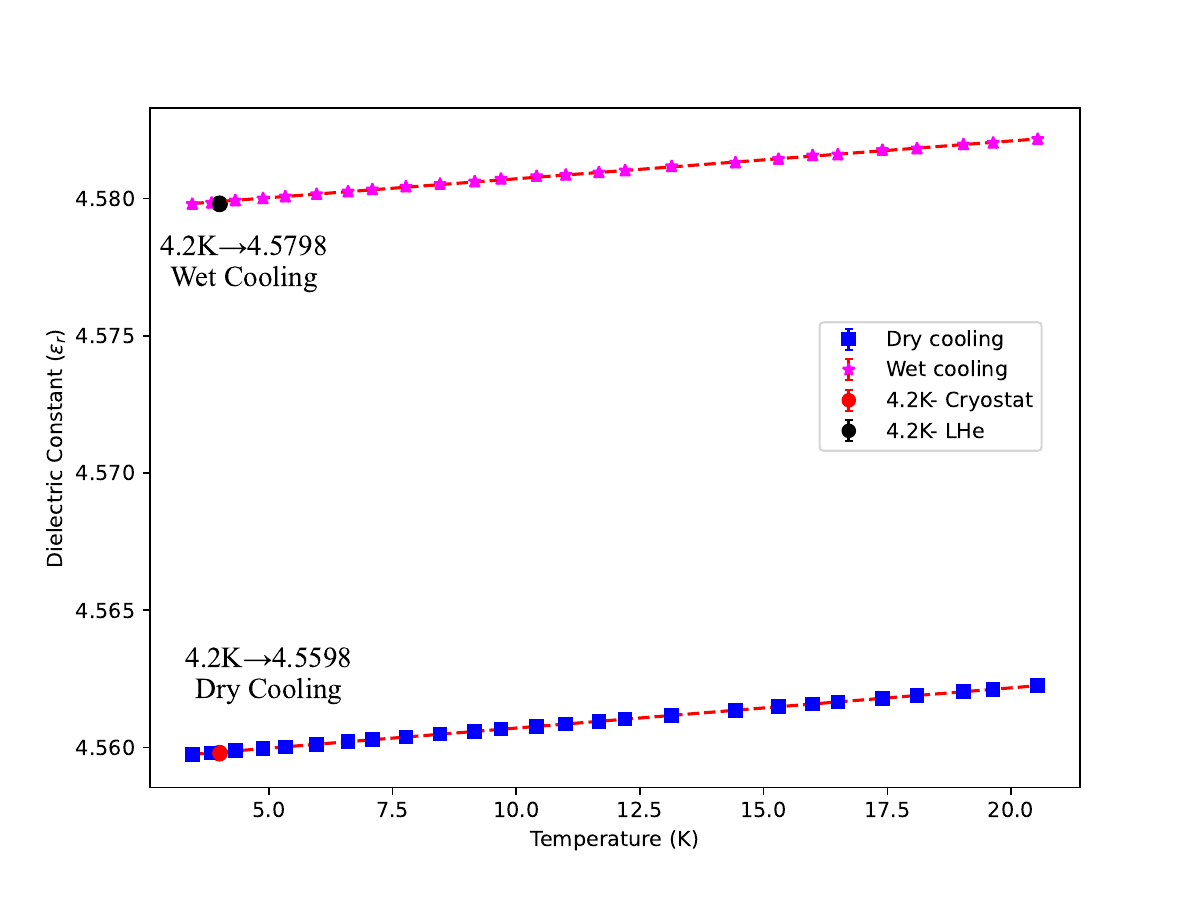}
         \label{fig:combined}
      \caption{ The dependence of dielectric constant on temperature for fused quartz. The calibration measurements were made in a dry cryostat, to compensate for the change in the electromagnetic environment seen by the capacitor. When we transition to a wet system for the Kapitza resistance measurements, we apply a linear shift to the data using the 4.2 K data points as the reference.}
    \label{fig:setup}

\end{figure}

\noindent 
To fabricate the samples, we followed the procedure that we outlined previously in \cite{babak}. We continue to fabricate capacitive thermometers measuring $20\times20$ mm using 500 $\pm$ 20 $\mu$m fused quartz substrates with 50 nm of gold (preceded by 3 nm Cr) on both sides on to which we deposit SiN deposition between 0 and 240 nm. Fused quartz (Inseto Ltd, UK) is used for the substrate as its dielectric constant continues to exhibit a dependence on temperature between 4 K and 22 K, as can be seen in Fig \ref{fig:setup}. The capacitance thermometers are calibrated using the same approach as previously discussed for liquid nitrogen \cite{babak}. The measurement set-up remains the same as in \cite{babak}, except that we reduced the thermalization time to 10 minutes upon immersion, considering the low temperature of the liquid helium. The samples are tested at six different power levels (ranging from 0.056 to 1.28 mW), corresponding to 2 to 10 mA with a current sweep in steps of 2 mA. 
\vskip -.6 cm

\newpage 
\noindent 
We assume that the volume of liquid helium in the dewar is considered large enough that applying current to the heater does not change the liquid helium temperature. Additionally,  it is worth noting that the heat provided by the LCR meter to the sample is on the order of nanowatts and can be safely neglected.\\
\vspace{-0.4cm}
\section{Measurement Results}\label{sec:IV}
\begin{figure}[ht]
    \centering
    \begin{subfigure}[b]{0.49\textwidth}
        \centering
        \includegraphics[width=\linewidth]{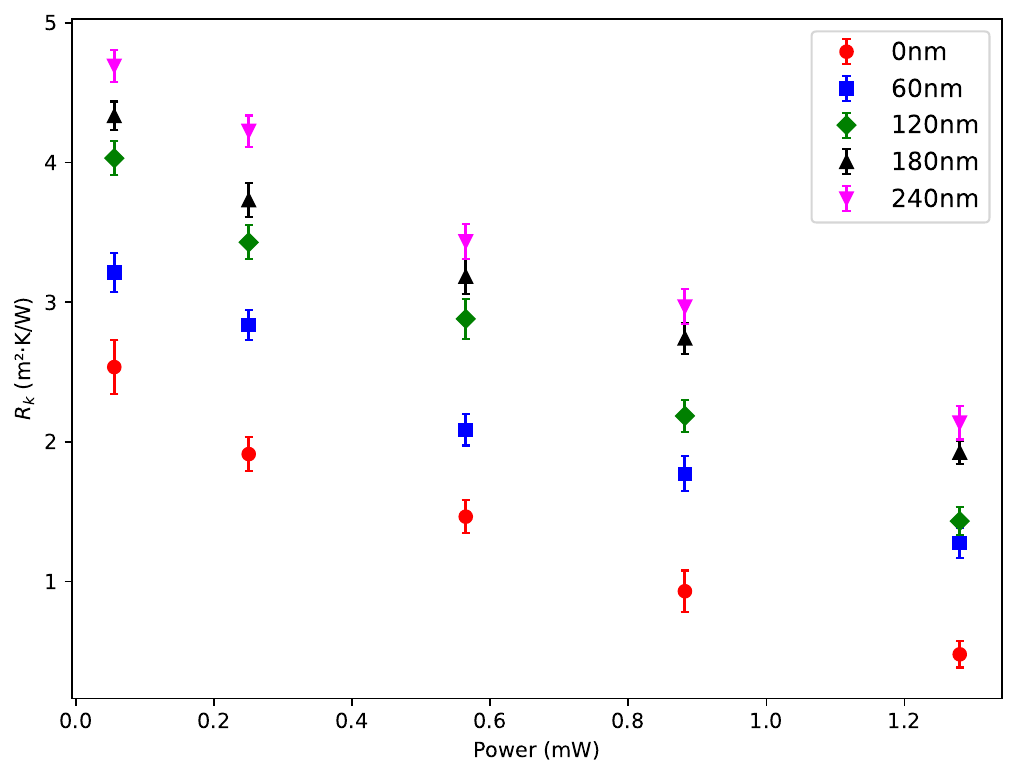}
        \caption{}
        \label{fig:KapitzaMeasurement}
    \end{subfigure}\hfill
    \hspace{-0.2cm}
    \begin{subfigure}[b]{0.49\textwidth} 
        \centering
        \includegraphics[width=\linewidth]{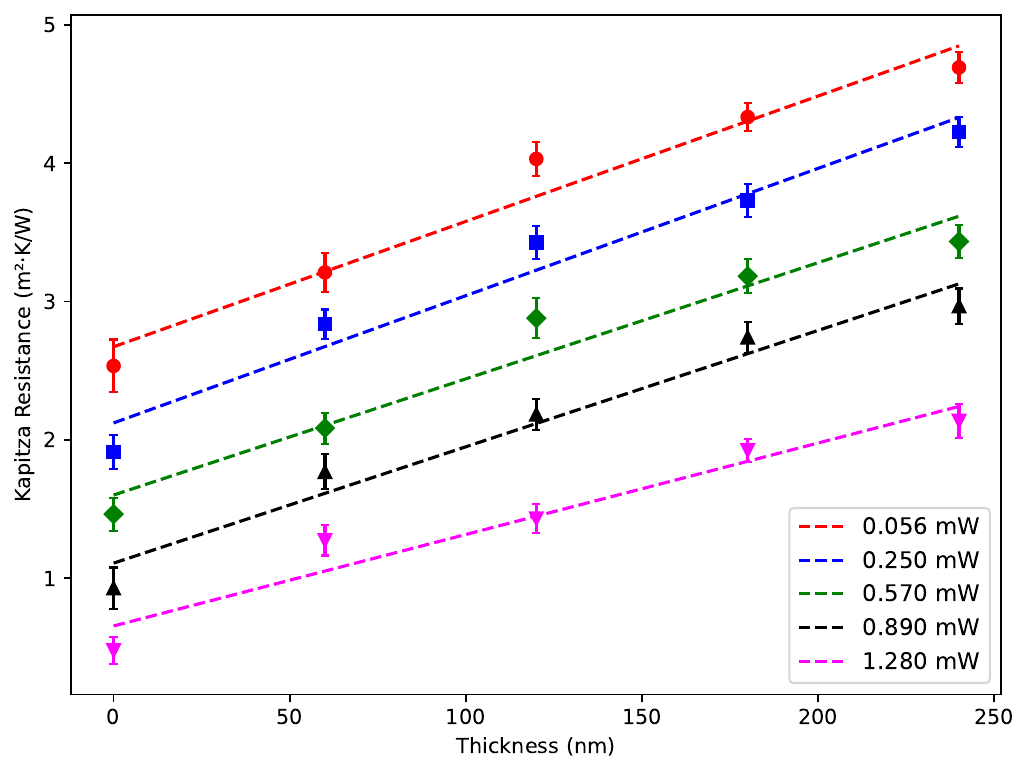}
        \caption{}
        \label{fig:thicknessline}
    \end{subfigure}
    \caption{(a) Kapitza resistance for different passivation layers at six different power levels ranging from 0.056 to 1.28 mW. (b) Kapitza resistance in terms of thickness for corresponding powers. The plot employs a linear regression fit, which models the connection between thickness and Kapitza Resistance.}
    \label{fig:kap}
\end{figure}

The Kapitza resistances for the 6 different powers and thicknesses of the passivation layer are shown in Fig. \ref{fig:kap}. 
\vspace{-0.3cm}

\vspace{-1em}
\noindent The illustrated data in Fig. \ref{fig:kap} shows that the thicker nitride passivation layer directly results in a higher Kapitza resistance at the interfacial area of the solid and the liquid helium. On average, for each additional nanometer of thickness, the Kapitza resistance increases by about 0.0365 m$^2$.K/W.\\
In our system, uncertainties arise in capacitance measurement and providing a linear fit, which subsequently affects the accuracy of the $T_2$ measurements. Additionally, in the calculation of $R_k$, these errors introduce further uncertainties due to Equation \ref{eq:1}.\\
\newpage 
\section{Conclusion} \label{sec:V}
In this paper, we have extended our earlier measurements to 4.2 K using liquid helium. We observed that depositing silicon nitride on a quartz substrate increases the Kapitza resistance by $\sim$ 0.0365 m$^2$.K/W  per nanometer. We suspect that the additional Kapitza resistance introduced by the passivation layer may explain some of the difficulty in removing heat from low noise amplifiers at cryogenic temperatures that has been reported in the literature \cite{13}. Consequently, we propose that one method to overcome the $R_K$ will be using a much thinner layer of SiN as an anti-reflective coating, and further work is underway to investigate this.

\newenvironment{acknowledgments}
  {\section*{Acknowledgments}}
  {}
\begin{acknowledgments}
This project received funding from the European Union’s Horizon 2020 research and innovation program under the Marie Skłodowska-Curie grant agreement No. 811312 for the project “Astro-Chemical Origins” (ACO) and UKRI ST/X006344/1.
We would like to thank Mark Seller, Dominic Mccullagh, Jesus Figueroa, and Darren Shepherd for facilitating the liquid helium test setup. Also,
we wish to acknowledge the support of the National Graphene Institute team, Dr. Lee Hague, Andrew Brook, Robert Howard, Matthew Whitelegg, and Dr. Kunal Lulla, offering suggestions in the fabrication process.

\end{acknowledgments}

\bibliography{sn-bibliography}

\end{document}